\documentclass[runningheads]{llncs}
\usepackage[T1]{fontenc}
\usepackage{graphicx}
\usepackage{graphicx}
\usepackage{verbatim}
\usepackage{multirow}
\usepackage{amsmath} 
\usepackage{array}
\usepackage{multirow}
\usepackage{tabularx}
\usepackage{caption}
\usepackage[T1]{fontenc}
\usepackage{hyperref}
\usepackage{amssymb}
\usepackage{latexsym}
\usepackage{url}
\usepackage{xcolor}
\usepackage{subcaption}
\usepackage{mathtools}
\usepackage{adjustbox}
\usepackage{float}
\newcolumntype{Y}{>{\centering\arraybackslash}X}
\usepackage{hyperref}
\usepackage{graphicx}
\usepackage{comment}
\begin{document}
\title{Leveraging Auto-Distillation and Generative Self-Supervised Learning in Residual Graph Transformers for Enhanced Recommender Systems}
%
%
\titlerunning{Leveraging Auto-Distillation and GSSL in RGT for Enhanced RS}

%
\author{Eya Mhedhbi\inst{1} \and
Youssef Mourchid\inst{2}\orcidID{0000-0003-4108-4557} \and
Alice Othmani\inst{3,1}\orcidID{0000-0002-3442-0578}}
\authorrunning{E. Mhedhbi et al.}
%
\institute{Declic AI Research \email {mhedhbiaya99@gmail.com} \and
CESI LINEACT Laboratory, UR 7527, Dijon, France
\email{ymourchid@cesi.fr}\\
\and
Université Paris-Est, LISSI, UPEC, 94400 Vitry sur Seine, France\\
Corresponding author: \email{alice.othmani@u-pec.fr}}
\maketitle              
\begin{abstract}
This paper introduces a cutting-edge method for enhancing recommender systems through the integration of generative self-supervised learning (SSL) with a Residual Graph Transformer. Our approach emphasizes the importance of superior data enhancement through the use of pertinent pretext tasks, automated through rationale-aware SSL to distill clear ways of how users and items interact. The Residual Graph Transformer incorporates a topology-aware transformer for global context and employs residual connections to improve graph representation learning. Additionally, an auto-distillation process refines self-supervised signals to uncover consistent collaborative rationales. Experimental evaluations on multiple datasets demonstrate that our approach consistently outperforms baseline methods. 

\keywords{Recommender Systems \and Generative Self-Supervised learning  \and Residual Graph Transformer \and Masked Autoencoder \and Representation Learning.}
\end{abstract}
\section{Introduction}
In today's age of information overload, recommender systems are essential for offering personalized suggestions for items that users might find appealing, such as products on e-commerce platforms\cite{1} and social media\cite{2}. These systems are integral to enhancing user engagement and facilitating discovery amid vast content repositories like Amazon, Alibaba, Facebook, YouTube, and TikTok. 

Collaborative filtering (CF) is widely employed within these systems, it generates suggestions by leveraging the preferences of similar users or items to recommend new items for a given user\cite{3}. In recent years, graphs have been extensively utilized in a variety of applications \cite{0}, with Graph Neural Networks (GNNs) enhancing collaborative filtering by modeling users and items as connected nodes. Models like PinSage \cite{4}, NGCF \cite{5}, and LightGCN \cite{6} use GCNs for better embeddings and accuracy. LightGCN \cite{6} focuses on structural signals, while GNNs face challenges with sparse and noisy data. 
Self-supervised learning (SSL), a powerful technique that trains models using unlabeled data, has emerged as a valuable approach in recommender systems by learning from unlabeled data. SSL methods are either contrastive, like SGL \cite{8}, which creates varied views of interaction graphs, or generative, like GFormer \cite{9}, which reconstructs masked interactions. Graph Contrastive Learning (GCL) is useful but struggles with noise. Traditional methods, such as hypergraph-based passing (HCCF) \cite{10} and node clustering (NCL) \cite{11}, address some issues but are also noise-prone.

Generative Learning \cite{02} in recommendation systems involves generating new data instances from existing ones. Self-supervised collaborative filtering leverages this approach through several advanced techniques: (1) Variational Autoencoders (VAE): Methods like CaD-VAE \cite{13} enhance modeling by integrating item content, optimizing hyperparameters, and using advanced frameworks to improve efficiency. (2) Masked Autoencoders: Techniques such as AutoCF \cite{14} and GFormer \cite{9} reconstruct data by focusing on significant user-item interactions. GFormer emphasizes autonomous edge detection and reconstruction using user and item representations. (3) Denoised Diffusion Models: Approaches like DiffRec \cite{15} use diffusion models to reduce noise from user ratings or latent variables, enhancing prediction accuracy and efficiency.

To overcome the limitations of the different approaches and in order to leverage the strengths of various techniques, we propose a novel recommender system called Residual Graph Transformer. It automatically extracts self-supervised signals, improves the graph autoencoder with adaptive methods, and uses Transformers to capture relationships and collaborative patterns for better recommendations. Auto-distillation is also used to improve the performance.
Our main contributions can be summarized as follows:
\begin{itemize}

\item We introduce automated ways to enhance data through self-supervised learning (SSL) and explain how these methods can make recommendations more robust.

\item We use graph transformer models with residual connections and light self-attention to identify consistent data patterns. Auto-distillation boosts recommendation performance, and a graph autoencoder enhances the representation of user-item interactions.

\item We evaluate our proposed model on multiple datasets and find it consistently outperforms other methods in various scenarios.

\end{itemize}

The organization of the rest of this paper is as follows. Section \ref{sec:proposedapproach} comprehensively details different components of our proposed model. Section \ref{sec:results} outlines the experimental setup and presents the results. Finally, concluding remarks are presented in Section \ref{sec:conclusion}.

\section{Proposed Approach}\label{sec:proposedapproach}

\subsection{Problem Formulation}
In this work, we consider a recommendation system with $I$ users $U = \{u_1, u_2, ..., u_I\}$ and $J$ items $P = \{p_1, p_2, ..., p_J\}$ . The user interactions are captured by an interaction matrix $A \in R^{I \times J}$, where $a_{i,j} = 1$  indicates an interaction between user $u_i$ and item $p_j$, and $a_{i,j} = 0$ otherwise. To convert the interaction matrix into an interaction graph for graph-based CF, we define a graph $G = \{V,E\}$, where $V = U \cup P$ represents the set of nodes, and $E = \{e_{i,j}|a_{i,j} = 1\}$ denotes the set of edges corresponding to user-item interactions. With these definitions, the graph-based CF can be represented as a prediction function for user-item interactions: $\hat{y}_{i,j} = f(G;\Theta)$, where $\hat{y}_{i,j}$ is the predicted score for the unknown interaction between user $u_i$ and item $p_j$, and $\Theta$ represents the model parameters.
\begin{figure}
\centering
\includegraphics[width=1.2\textwidth]{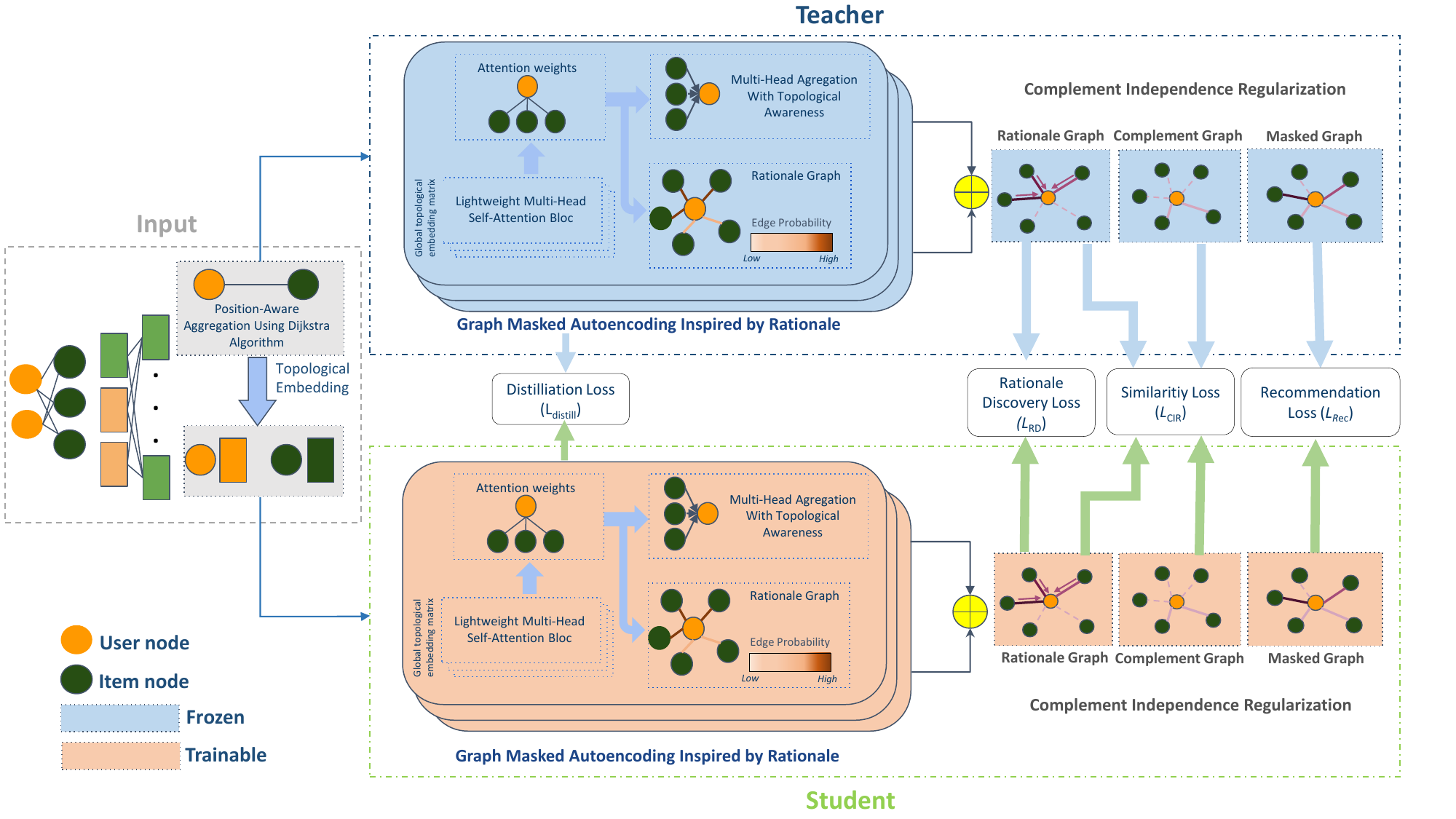}
\caption{Overall framework illustration of our proposed approach. The figure is better seen zoomed in the digital version of the document.} \label{fig:my_label}
\end{figure}
\subsection{Integration of Global Topology Information}
Inspired by the efficiency of position-aware graph neural networks(P-GNNs)\cite{18} in capturing comprehensive relational information on a global scale, we initially select a subset of anchor nodes $V_A \subset V$ from the user-item interaction graph $G = \{V, E\}$. We then compute the distance between a target node \(v_k\) and the anchor sets \(v_a\) using Dijkstra’s algorithm. Given the calculated distances, we compute the correlation weight, denoted as $\omega_{k,a}$, for each pair of target-anchor nodes $(k, a)$. The correlation weight is defined as follows:

\begin{equation}
\omega_{k,a} = 
\begin{cases} 
\frac{1}{d_{k,a}+1} & \text{if } d_{k,a} \leq q \\
0 & \text{otherwise}.
\end{cases}
\end{equation}

q is the upper limit for the correlation weight between any target and anchor node. This method ensures that only nodes within a certain distance contribute to the target node’s embedding, effectively capturing the positional relationships within the graph. Dijkstra’s algorithm is used for a more accurate representation of these relationships, resulting in more informative node embeddings. This holds true even when edge weights are binary, Dijkstra’s algorithm can provide a more nuanced understanding of the graph structure, which can lead to improved performance. The correlation weights of the nodes are normalized to a range of [0, 1]. Using the weight $\omega_{k,a}$, the target node embedding is refined by considering the correlation weights between the target node $k$ and each anchor node $v_a \in V_A$: $\tilde{h}^l_k = \sum_{v_a \in V_A} W^l \cdot \omega_{k,a} \cdot [\tilde{h}^{l-1}_k || \tilde{h}^{l-1}_a] / |V_A|
$

Here, $\tilde{h}^l_k$ and $\tilde{h}^{l-1}_k \in R^d$ denote the embeddings of node $v_k$ in the $l$-th and $(l-1)$-th graph propagation layer, respectively. To capture the global topological context based on anchor nodes, $H^l_T$ is defined as the global topological embedding matrix in the $l$-th layer. $W^l \in R^{d \times 2d}$ represents a learnable transformation matrix. After $L$ graph information propagation steps, the embeddings in $\tilde{H}^L$ retain the high-order topological information. This information is then injected into the id-corresponding embeddings to produce the topological embeddings:
\begin{equation}
\bar{H} = TE(H; \{W^l_T\})
\end{equation}
In this manner, our parameterized rationale generator can capture global collaborative relationships and uncover informative interaction patterns between users and items for SSL augmentation.
\subsection{Residual Graph Transformer-Based Collaborative Rationale Discovery}

Rational discovery helps to find useful patterns in how users interact with items. These patterns can help improve recommendation systems in changing environments, even when there are few labels for supervision. We propose an advanced method for extracting these rationales using our Graph Transformer model with residual connections. Our method is inspired by the core self-attention mechanisms of the Transformer architecture, focusing on invariant user preference information for robust generative self-supervision. This approach mitigates noise and biases in observational behavior data, enhancing recommendation performance. The core of our method is a rationale discovery module within a graph transformer framework, encoding node-wise relations as selected rationales and incorporating the positional information of user and item nodes. We leverage global topology-aware node embeddings, $\bar{H}$, in a multi-head self-attention mechanism for the rationalization process.
The correlation between nodes $v_k$ and $v_{k'}$ for the $h$-th attention head is computed as:
\begin{equation}
\alpha^h_{k,k'} = \frac{\exp(\tilde{\alpha}^h_{k,k'})}{\sum_{k'} \exp(\tilde{\alpha}^h_{k,k'})} ; \tilde{\alpha}^h_{k,k'} = \frac{(W^h_Q \cdot \bar{h}_k)^T \cdot (W^h_K \cdot \bar{h}_{k'})}{\sqrt{d/H}}
\end{equation}
Here, $\alpha^h_{k,k'}$ represents the normalized attention score between nodes $v_k$ and $v_{k'}$ for the $h$-th attention head, and $\tilde{\alpha}^h_{k,k'}$ is the raw attention score before normalization. $W^h_Q$ and $W^h_K$ in $\mathbb{R}^{d_H \times d}$ are the transformations used to derive the query and key embeddings for calculating the attention scores. As the attention scores encoded by our graph transformer reflect the strength of node-wise dependencies, we aggregate the multi-head scores to obtain the probability scores, $p((v_k,v_{k'})|G)$, of graph edges, such as ($v_k - v_{k'})$, being selected as rationales. These rationales represent a subset of crucial user-item interaction patterns that most effectively highlight the user preference learning process with consistent representations, as shown:

\begin{equation}
\bar{p}((v_k,v_{k'})|G) = \frac{\bar{\alpha}_{k,k'}}{\sum_{(v_k,v_{k'}) \in E} \bar{\alpha}_{k,k'}} ; \bar{\alpha}_{k,k'} = \frac{\sum_{h=1}^H \alpha^h_{k,k'}}{H}
\end{equation}
To sample a rationale estimated by our topology-aware graph transformer, we individually select $\rho_R \cdot |E|$ edges from the edge set $E$ based on their probability scores, $p((v_k,v_{k'})|G)$. Here, the hyperparameter $\rho_R \in R$ determines the proportion of important edges chosen for rationalization. To improve our model, we added a residual operation to our residual graph transformer. This upgrade builds on the basic graph transformer by repeatedly applying it and combining the results through a residual link. This helps maintain the flow of gradients during training, making learning more efficient and improving the model's performance. This residual modification significantly boosts the model's ability to find and understand important patterns in user-item interactions, enhancing recommendation effectiveness in changing environments with limited supervised data. In addition to the above, we have also incorporated a streamlined version of self-attention, into our model. This version employs linear layers for transforming the input vectors, which facilitates easier initialization and improved gradient propagation. An output projection layer has been integrated to transform the resulting embeddings back to the original dimensionality. This makes the Light Self-Attention particularly suitable for large datasets and deep architectures, while being more computationally and memory efficient.
\subsection{Self-Augmentation with Rationale Awareness}

\textbf{Integrating Rationales into Graph Masked Autoencoding}. The approach uses a graph-masked autoencoder for self-augmentation, leveraging collaborative rationales. The rationale-aware mask autoencoder masks identified rationales from the interaction graph, allowing for effective reconstruction. The masked graph structure $G_M = \{V, E_M\}$ is created by inverting the rationale scores. The masked graph maintains a higher edge density than the rationale graph, ensuring only the most significant rationale edges are removed, promoting noise-resistant autoencoding. The process begins with the following operation:
\begin{equation}
S = GT(G_M, TE(\bar{S}_L)); \bar{s}_l = \sum_{(v_k,v_{k'}) \in E_M} \beta_{k,k'} \cdot \bar{s}_{l1_{k'}}
\end{equation}
Here, $GT(·)$ and $TE(·)$ represent the graph transformer network and the topological information encoder, respectively. $\bar{s}_l$ is the sum of the product of $\beta_{k,k'}$ and $\bar{s}_{l1_{k'}}$ for all edges in $E_M$. This equation is used to enhance the initial embeddings with $L$-order local node embeddings $\bar{S}_L$, encoded from LightGCN. The embeddings $S$ are then used to train the reconstruction of the masked user-item interactions. The final embeddings $S \in R^{(I+J) \times d}$ in the autoencoder are utilized for training the reconstruction of the masked user-item interactions, and for computing the masked autoencoder loss $L_{MAE}$, which is defined as follows:
\begin{equation}
L_{MAE} = \sum_{(v_k, v_k') \in E \setminus E_M} -\tilde{y}_{k,k'} ;   \tilde{y}_{k,k'} = s_k^\top s_{k'}
\end{equation}
This loss is specifically designed for reconstructing masked interaction patterns, $\tilde{y}_{k,k'}$ represents the predicted scores for the edge $(v_k, v_{k'})$ on graph G. The reconstruction process is enhanced by encoding local node embeddings from LightGCN into the initial embeddings, ensuring effective reconstruction of key interaction patterns adaptable to downstream recommendation tasks. This approach helps prevent noise in the generative self-supervised learning process.\\
\textbf{Modeling Complement Independence}.To make the rationale discovery process more independent and reduce redundancy, we use a learning technique called contrastive regularization. This minimizes the mutual information between the rationale graph\(G_R\) and a complement graph \(G_C\). We create the complement graph\(G_C\) by sampling edges with a low rate \(\rho_C\) << \(\rho_M\) to focus on noisy edges. The edges of the complement graph (E\_C) are generated as follows:
\begin{equation} 
E_C \sim p_C(E_C|G) = \rho_M(E_C|G); |E_C| = \rho_C \cdot |E| \end{equation}
This is similar to graph masking but with a smaller sampling rate. Then, we used the loss function \(\mathcal{L}_{CIR}\), to reduce the similarities between the rationale graph \(G_R\) and the complement graph \(G_C\) in high-order representations, which can be defined as follows:
\begin{equation}
\mathcal{L}_{CIR} = \log \sum_{v_k \in V} \exp \left( \frac{\cos(e^R_k, e^C_k)}{\tau} \right)
\end{equation}
Here, \(e^R_k\) and \(e^C_k\) are the embeddings of node \(k\) in the rationale graph and complement graph, respectively. The cosine similarity between these embeddings is divided by a temperature coefficient \(\tau\), and the exponential sum of these values across all nodes is taken. This loss function ensures that the embeddings of the rationale graph and the complement graph are pushed apart, enhancing the model's ability to separate useful information from noise.\\
\textbf{SSL-Augmented Model Optimization}. During training, we use embeddings \(S \in R^{(I+J) \times d}\) to predict interactions for training the recommender system. The point-wise loss function \(L_{Rec}\) is minimized as follows:

\begin{equation}
L_{Rec} = \sum_{i,j=1}^{A} -\log \left( \frac{\exp(s_i^T s_j)}{\sum_{p' \in P} \exp(s_i^T s_{p'})} \right)
\end{equation}
Our Model maximizes predictions for all positive user-item interactions and contrasts by minimizing predictions for negative interactions. During testing, we substitute the masked graph \(G_M\)  with the observed interaction graph \(G\) and predict user \(u_i\)'s relation to item \(p_j\) as \(\hat{y}_{i,j} = s_i^T s_j\).
\subsection{Auto Distillation Graph}
Auto-distillation improves model training by using a single model as both teacher and student. After steps like negative sampling and embedding generation, the model’s embeddings (user, item, and graph structure) are used to train a second model, the distillation model. The distillation model learns to mimic the main model’s outputs, guided by a loss function, Mean Squared Error (MSE), which reduces differences between the outputs of the main model and the distillation model. This process can be expressed as:
\begin{align}
L_{\text{distill}} = & \text{MSE}(h^{\text{student}}_{u}, h^{\text{teacher}}_{u}) + \text{MSE}(h^{\text{student}}_{i}, h^{\text{teacher}}_{i}) \nonumber \\
& + \text{MSE}(h^{\text{student}}_{c}, h^{\text{teacher}}_{c}) + \text{MSE}(h^{\text{student}}_{s}, h^{\text{teacher}}_{s})
\end{align}
where \(h^{student}_{u}\) and \(h^{teacher}_{u}\) represent the user embeddings generated by the student and teacher models, respectively, \(h^{student}_{i}\) and \(h^{teacher}_{i}\) represent the item embeddings, \(h^{student}_{c}\) and \(h^{teacher}_{c}\) represent the contrastive list embeddings, and \(h^{student}_{s}\) and \(h^{teacher}_{s}\) represent the subgraph list embeddings.

\subsection{Our Objective Loss}
The BPR loss, $L_{RD}$, guides rationale discovery by optimizing the model's objective for the downstream task:
\begin{equation}
L_{RD} = \sum_{(u_i, p^+_j, p^-_j)} - \log \text{sigm}(\bar{y}_{i,j+} - \bar{y}_{i,j-})
\end{equation}
Here, a triplet of user $u_i$ and items $p^+_j, p^-_j$ is sampled such that $(u_i, p^+_j) \in E$ and $(u_i, p^-_j) \notin E$. The sigmoid function, $\text{sigm}(\cdot)$, models the probability that the positive interaction $p^+_j$ is preferred over $p^-_j$.\\
In our model, the total loss function \(L\) integrates several key objectives:  \(L_{\text{Rec}}\), \(L_{\text{RCS}}\), \(L_{\text{RD}}\), \(L_{\text{CIR}}\), and the Frobenius norm regularization \(||\Theta||_F^2\). Additionally, the distillation loss \(L_{\text{distill}}\) refines the model's internal representations over training epochs, enhancing the accuracy of recommendations. The overall objective \(L\) to be minimized during training consolidates these components:
\begin{equation}
L = L_{\text{Rec}} + \lambda_{\text{distill}} L_{\text{distill}} + \lambda_1 \cdot L_{\text{RD}} + \lambda_2 \cdot L_{\text{CIR}} + \lambda_3 \cdot ||\Theta||_F^2
\end{equation}
Here, \(\lambda_1\), \(\lambda_2\), \(\lambda_3\), and \(\lambda_{\text{distill}}\) are hyperparameters balancing the impact of each loss term. The Frobenius norm regularization \(||\Theta||_F^2\) helps keep the model stable by limiting parameter values. This approach balances reconstruction, rationale discovery, and regularization for better optimization in recommendation systems.
\section{Experiments and results}\label{sec:results}

\subsection{Experimental Settings}
\subsubsection{Datasets.}We use three real-world datasets to assess the performance of our proposed model. These include the Yelp dataset, which focuses on recommending business venues from the renowned Yelp platform; the Ifashion dataset, a collection of fashion outfit data gathered by Alibaba, and the LastFM dataset, which records user interactions within music applications and internet radio platforms. The characteristics and statistics of these datasets are detailed in Table \ref{tab:data}.
\begin{table}
\centering
  \caption{Statistics of the experimental datasets}
\begin{adjustbox}{max width=1\textwidth}
\begin{tabular}{c c c c c} 
    \hline
    Dataset & Users & Items & Interactions & Density \\
    \hline
    Yelp & 42,712 & 26,822 & 182,357 & $1.6e^{-4}$ \\
    Ifashion & 31,668 & 38,048 & 618,629 & $5.1e^{-4}$ \\
    LastFM & 1,889 & 15,376 & 51,987 & $1.8e^{-3}$ \\
    \hline
\end{tabular}
\end{adjustbox}
\label{tab:data}
\end{table}
\subsubsection{Evaluation Protocols.}For each dataset, interactions are split into training, validation, and test sets with a division ratio of 70\%, 5\%, and 25\%, respectively. To evaluate the effectiveness of our proposed model, we employ a comprehensive all-rank evaluation method that mitigates potential biases from negative sampling. The performance of the models is assessed using two key metrics: Recall Ratio (Recall@K) and Normalized Discounted Cumulative Gain (NDCG@K) for \( K = 10, 20, 40 \). 
NDCG is calculated using the formula:
\begin{equation}
NDCG@k = \frac{DCG@k}{IDCG@k}
\end{equation}
Where \(DCG@k\) is the Discounted Cumulative Gain at position k, calculated by summing the gains of the recommendations up to position k and giving more weight to higher-ranked recommendations, and \(IDCG@k\) is the Ideal Discounted Cumulative Gain at position k, representing the optimal possible DCG when the recommendations are perfectly ranked. 

\subsection{Hyperparameter Settings}
We used Weights \& Biases (W\&B) \footnote{\url{https://wandb.ai/site}} for hyperparameter tuning to maximize Recall and NDCG values. Table~\ref{tab:param} presents the best hyperparameter combinations for the different datasets.

\begin{table}
\centering
  \caption{The best hyperparameter combinations used across various datasets in our model.}
\begin{adjustbox}{max width=1\textwidth}
\begin{tabular}{c c c c c c c c c c c c c c } 
    \hline
    Dataset & anchor\_set & reg & lr & ssl\_reg & GCN & B2 & PNN& Batch & ctra & GT & GTW & Head & latdim \\
    \hline
    Yelp & 16 & 0.0001 & 0.001 & 0.5 & 3 & 0.5&2 &  4096 & 0.005 & 2 & 0.05 & 2 &  64 \\
    Ifashion & 64 & 0.00001 & 0.001 & 1.5 & 3 & 1 & 1 & 4096 &  0.0005 & 1 & 0.05 & 2 & 32 \\
    LastFM & 32 & 0.0001 & 0.001 & 0.5 & 1 & 1 & 2 &  4096 & 0.005 & 1 & 0.1 & 8 & 64 \\
    \hline
\end{tabular}
\end{adjustbox}
\label{tab:param}
\end{table}

\subsection{Ablation Study}
\begin{figure}
    \centering
    \begin{subfigure}[b]{0.45\textwidth}
        \includegraphics[width=\textwidth]{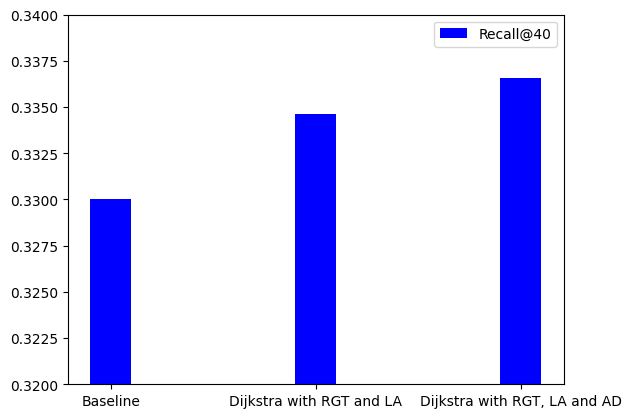}
        \caption{LastFM Dataset - Recall@40}
        \label{fig:sub1}
    \end{subfigure}
    \hfill
    \begin{subfigure}[b]{0.45\textwidth}
        \includegraphics[width=\textwidth]{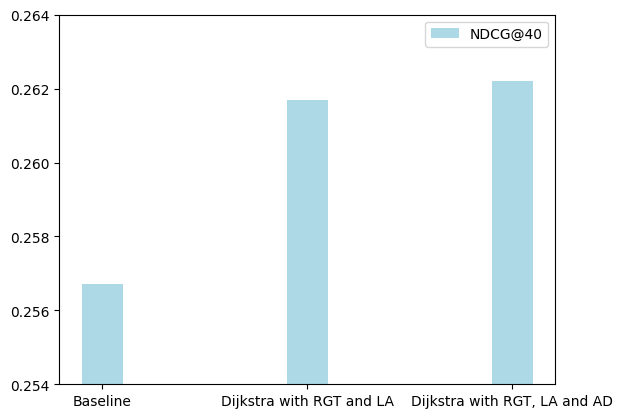}
        \caption{LastFM Dataset - NDCG@40}
        \label{fig:sub2}
    \end{subfigure}
    
    \caption{Effect of different components of our proposed models on LastFM dataset in terms of Recall@40 and NDCG@40}
    \label{fig:figures_globales}
\end{figure}

To validate the effectiveness of our contributions, we conducted an ablation study comparing three versions of our model using the LastFM dataset. The first version, Graph Transformer, served as our baseline model. We then added Dijkstra's algorithm and the Residual Graph Transformer(RGT) with light attention(LA) in the second version, followed by the application of auto-distillation in the third version(AD). As shown in Fig.~\ref{fig:figures_globales}, each enhancement led to progressive improvements in Recall@40 and NDCG@40 metrics, underscoring the effectiveness of our proposed methods.

\begin{figure}
    \centering
    \begin{subfigure}[b]{0.45\textwidth}
        \includegraphics[width=\textwidth]{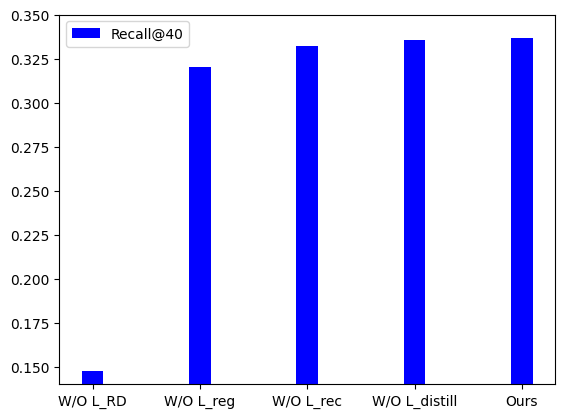}
        \caption{LastFM Dataset - Recall@40}
        \label{fig:sub1}
    \end{subfigure}
    \hfill
    \begin{subfigure}[b]{0.45\textwidth}
        \includegraphics[width=\textwidth]{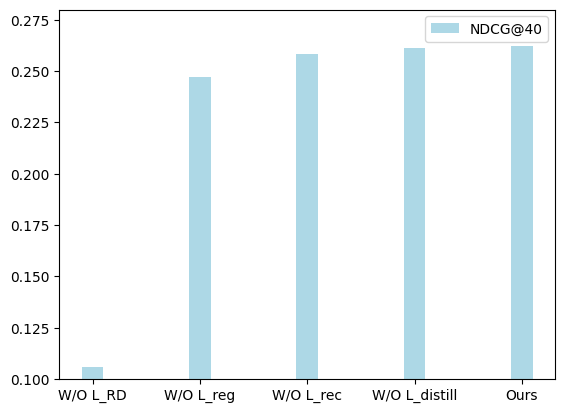}
        \caption{LastFM Dataset - NDCG@40}
        \label{fig:sub2}
    \end{subfigure}
    
    \caption{Effect of different losses used in our model on the Recall@40 and NDCG@40 metrics for the LastFM dataset}
    \label{fig:losses}
\end{figure}
Additionally, as mentioned in Fig.~\ref{fig:losses} ,we assess the impact of individual loss compo-nents on model performance, we conducted an ablation study by sequentially removing each component and measuring changes in Recall@40 and NDCG@40. Results show that the L\_RD component is crucial for guiding rationale discovery and optimizing the model, with its removal leading to significant drops in both metrics. The L\_rec component is also vital, as its absence results in a major performance decline. The L\_distill component, while less critical, still provides a positive but modest contribution. The L\_reg component is essential for preventing overfitting, as indicated by the performance decrease when it is removed. Overall, the full model with all loss components included demonstrates the highest performance, underscoring the effectiveness of the combined loss strategy.
\begin{table}[!ht]
\centering
\caption{Performance comparison between our proposed approach and all baselines on Ifashion, Yelp, LastFM datasets.
\textcolor{red}{}}
\begin{adjustbox}{max width=1.25\textwidth}
\begin{tabular}{c c c c c c c c c c c c c c c c} 

\hline
Datasets & Metric & BiasMF & NCF & AutoRec & PinSage & NGCF & GCCF & LightGCN & EGLN & SLRec & NCL & HCCF & SGL & GFormer & Ours \\
\hline
Ref & \_ & \cite{19} & \cite{3} & \cite{20} & \cite{4} & \cite{5} & \cite{21} & \cite{6} & \cite{17} & \cite{16} & \cite{11} & \cite{10} & \cite{8} & \cite{9} & \_ \\
\hline
\multirow{6}{*}{Yelp} & Recall@10 & 0.0122 & 0.0166 & 0.0230 & 0.0278 & 0.0438 & 0.0484 & 0.0422 & 0.0458 & 0.0418 & 0.0493 & 0.0518 & 0.0522 & 0.0562 & \textbf{0.0583} \\
& NDCG@10 & 0.0070 & 0.0101 & 0.0133 & 0.0171 & 0.0269 & 0.0296 & 0.0254 & 0.0278 & 0.0258 & 0.0301 & 0.0318 & 0.0319 & 0.0350 & \textbf{0.0363} \\
& Recall@20 & 0.0198 & 0.0292 & 0.0410 & 0.0454 & 0.0678 & 0.0754 & 0.0761 & 0.0726 & 0.0650 & 0.0806 & 0.0789 & 0.0815 & 0.0878 & \textbf{0.0911} \\
& NDCG@20 & 0.0090 & 0.0138 & 0.0186 & 0.0224 & 0.0340 & 0.0378 & 0.0373 & 0.0360 & 0.0327 & 0.0402 & 0.0391 & 0.0410 & 0.0442 & \textbf{0.0458} \\
& Recall@40 & 0.0303 & 0.0442 & 0.0678 & 0.0712 & 0.1047 & 0.1163 & 0.1031 & 0.1121 & 0.1026 & 0.1192 & 0.1244 & 0.1249 & 0.1328 & \textbf{0.1377} \\
& NDCG@40 & 0.0117 & 0.0167 & 0.0253 & 0.0287 & 0.0430 & 0.0475 & 0.0413 & 0.0456 & 0.0418 & 0.0485 & 0.0510 & 0.0517 & 0.0551 & \textbf{0.0572} \\
\hline
\multirow{6}{*}{Ifashion} & Recall@10 & 0.0302 & 0.0268 & 0.0309 & 0.0291 & 0.0375 & 0.0373 & 0.0437 & 0.0473 & 0.0373 & 0.0474 & 0.0489 & 0.0512 & 0.0542 & \textbf{0.0555} \\
& NDCG@10 & 0.0281 & 0.0253 & 0.0264 & 0.0276 & 0.0350 & 0.0352 & 0.0416 & 0.0438 & 0.0353 & 0.0446 & 0.0464 & 0.0487 & 0.0520 & \textbf{0.0526} \\
& Recall@20 & 0.0523 & 0.0451 & 0.0537 & 0.0505 & 0.0636 & 0.0639 & 0.0751 & 0.0787 & 0.0633 & 0.0797 & 0.0815 & 0.0845 & 0.0894 & \textbf{0.0910} \\
& NDCG@20 & 0.0360 & 0.0306 & 0.0351 & 0.0352 & 0.0442 & 0.0445 & 0.0528 & 0.0549 & 0.0444 & 0.0558 & 0.0578 & 0.0603 & 0.0635 & \textbf{0.0650} \\
& Recall@40 & 0.0858 & 0.0785 & 0.0921 & 0.0851 & 0.1062 & 0.1047 & 0.1207 & 0.1277 & 0.1043 & 0.1283 & 0.1306 & 0.1354 & 0.1424 & \textbf{0.1435} \\
& NDCG@40 & 0.0474 & 0.0423 & 0.0483 & 0.0470 & 0.0585 & 0.0584 & 0.0677 & 0.0715 & 0.0582 & 0.0723 & 0.0744 & 0.0773 & 0.0818 & \textbf{0.0826} \\
\hline
\multirow{6}{*}{LastFM} & Recall@10 & 0.0609 & 0.0574 & 0.0543 & 0.0899 & 0.1257 & 0.1230 & 0.1490 & 0.1133 & 0.1175 & 0.1491 & 0.1502 & 0.1496 & 0.1573 & \textbf{0.1610} \\
& NDCG@10 & 0.0696 & 0.0645 & 0.0599 & 0.1046 & 0.1489 & 0.1452 & 0.1739 & 0.1263 & 0.1384 & 0.1745 & 0.1773 & 0.1775 & 0.1831 & \textbf{0.1886} \\
& Recall@20 & 0.0980 & 0.0956 & 0.0887 & 0.1343 & 0.1918 & 0.1816 & 0.2188 & 0.1823 & 0.1747 & 0.2196 & 0.2210 & 0.2236 & 0.2352 & \textbf{0.2385} \\
& NDCG@20 & 0.0860 & 0.0800 & 0.0769 & 0.1229 & 0.1759 & 0.1681 & 0.2018 & 0.1557 & 0.1613 & 0.2021 & 0.2047 & 0.2070 & 0.2145 & \textbf{0.2180} \\
& Recall@40 & 0.1450 & 0.1439 & 0.1550 & 0.1990 & 0.2794 & 0.2649 & 0.3156 & 0.2747 & 0.2533 & 0.3130 & 0.3184 & 0.3194 & 0.3300 & \textbf{0.3366} \\
& NDCG@40 & 0.1067 & 0.1055 & 0.1031 & 0.1515 & 0.2146 & 0.2049 & 0.2444 & 0.1966 & 0.1960 & 0.2437 & 0.2458 & 0.2498 & 0.2567 & \textbf{0.2622} \\
\hline
\end{tabular}
\end{adjustbox}
\label{tab:my_label}
\end{table}

\subsection{Comparison with existing methods}
We present the performance comparison between our proposed approach and several baseline methods on the Ifashion, Yelp, and LastFM datasets in Table~\ref{tab:my_label}. Our approach consistently surpasses the baselines across all datasets and metrics. For instance, on the Yelp dataset, our approach achieves the highest Recall@10 (0.0583) and NDCG@10 (0.0363), significantly surpassing the other models. Similar trends are observed in the Ifashion and LastFM datasets, where our model attains the best results in Recall and NDCG at various K values, with notable improvements in Recall@40 and NDCG@40. These results demonstrate the efficacy of our model, particularly the residual graph transformer and the automatic distillation strategies, which contribute to superior performance by efficient mining and leveraging collaborative rationales. Among these components, the topological graph position encoding and the multi-head self-attention mechanism are crucial, as they capture global context and enhance the understanding of user-item interactions

\section{Conclusion}
\label{sec:conclusion}
This paper highlights how discovering meaningful user-item interaction patterns can improve collaborative filtering techniques through self-supervised learning. For that, we propose in this work a recommendation system that uses a residual graph transformer to automatically find hidden signals and coherent patterns. Our method includes topological graph position encoding for a global context and uses automatic distillation strategies to boost performance. Experiments show that our model outperforms several literature approaches across multiple datasets. In future work, we plan to explore advanced data augmentation techniques and expand our model to handle larger datasets and more diverse recommendation scenarios.

\section*{Acknowledgement}
This work was supported by Declic, a deep tech startup. Declic is a social network for outings, meetings, and events, fostering genuine connections and shared activities in the real world (https://declic.net/)

%
%
\bibliographystyle{splncs04}
\bibliography{refs}

\end{document}